\documentclass[oneside, a4paper, onecolumn, 11pt]{article}
\usepackage[left=2.5cm,top=2.5cm,bottom=2.5cm,right=2.5cm]{geometry}
\usepackage{lineno}
\usepackage{amsmath}
\usepackage{graphicx}
\usepackage{fancyhdr}
\usepackage{calc}
\usepackage{amssymb}
\usepackage{titlesec}
\usepackage{color}
\usepackage{booktabs}
\usepackage{float}
\usepackage{placeins}
\usepackage{enumitem}
\usepackage[super,sort&compress,comma]{natbib}
\usepackage{url}
\usepackage{setspace}
\usepackage{amsfonts}
\usepackage{commath}
\usepackage[innercaption]{sidecap}
\usepackage[framemethod=TikZ]{mdframed}
\usepackage{parskip}
\usepackage{pifont}
\usepackage[labelfont=bf]{caption}
\usepackage{hyperref}
\hypersetup{
    colorlinks=true,
    linkcolor=black,
    filecolor=gray,
    urlcolor=blue,
    citecolor=black,
}
\expandafter\def\expandafter\normalsize\expandafter{%
    \normalsize
    \setlength\abovedisplayskip{0pt}
    \setlength\belowdisplayskip{5pt}
    \setlength\abovedisplayshortskip{0pt}
    \setlength\belowdisplayshortskip{5pt}
}

\definecolor{Gray}{gray}{0.75}
\definecolor{changecolour}{rgb}{0, 0, 0.8}

\newmdenv[backgroundcolor=Gray, leftmargin = 0pt, rightmargin = 0pt, linewidth = 0pt, roundcorner = 2 pt, innerleftmargin=5pt, innerrightmargin=5pt, innertopmargin=5pt, innerbottommargin=5pt]{Frame}

\let\oldequation\equation
\let\oldendequation\endequation
\renewenvironment{equation}{\linenomathNonumbers\oldequation}{\oldendequation\endlinenomath}

\let\oldalign\align
\let\oldendalign\endalign
\renewenvironment{align}{\linenomathNonumbers\oldalign}{\oldendalign\endlinenomath}

\let\oldgather\gather
\let\oldendgather\endgather
\renewenvironment{gather}{\linenomathNonumbers\oldgather}{\oldendgather\endlinenomath}

\begin{document}

\begin{center}
    {\LARGE \textbf{Modest Algorithmic Mediation can Maximize Topical Diversity in Hybrid Human-AI Systems}}

    \vspace{2mm}
    Dini Wang$^{1,2}$ and Ho-Chun Herbert Chang$^{1,*}$
\end{center}

\small{
    \begin{enumerate}
        \item[$^1$]\textit{Program of Quantitative Social Science, Dartmouth College, Hanover, NH, USA}
        \item[$^2$]\textit{Department of Mathematics, Dartmouth College, Hanover, NH, USA}
        \item[$^*$]\textit{Correspondence: Herbert.Chang@dartmouth.edu}
    \end{enumerate}
}

\vspace{5mm}

\noindent\textbf{Abstract.} In the artificial intelligence (AI) era, the rise of algorithmic feeds has fundamentally transformed information diffusion on social media. 
While early platforms organized visibility through explicit social networks, contemporary systems mediate exposure through intelligent recommender algorithms that personalize attention. 
This paper examines how the social network and algorithmic architecture jointly shape the diversity of information sharing. 
Analysis of 18,076 users active throughout 2014--2018 shows that the topical diversity of sharing rose and then plateaued after the introduction of algorithmic ranking in 2016 while its inequality across users emerged alongside it. 
To this end, we introduce a hybrid human–AI information diffusion model in which information exposure is governed by a parameterized mixture of social propagation through the user-following network and algorithmic recommendation.
Both qualitative analysis and simulations show that the effect of algorithmic mediation is non-monotonic. Modest mediation can raise average diversity and reduce inequality relative to a purely network-driven baseline, whereas strong mediation reduces diversity and concentrates it among fewer users. Fitting the model to four years of data yields a mediation share that increases from zero before 2016 to approximately 0.50 by 2018, a level that exceeds the compensation point of equality while remaining within the diversity-enhancing range. These results identify the conditions under which recommendation broadens rather than narrows exposure and provide a unified framework for information diffusion in hybrid human-AI systems.

\vspace{3mm}

\noindent\textbf{Keywords:} algorithmic diffusion $|$ social network $|$ recommender systems $|$ exposure diversity $|$ information equality

\vspace{5mm}
\section*{Introduction}

The diffusion of information on social media was once understood primarily as a network process: messages spread through follower ties, friendship links, and repeated interpersonal sharing \citep{rogers2003diffusion,granovetter1973strength,watts2002simple,kempe2003maximizing,lerman2010information,bakshy2012role,vosoughi2018spread}. Early platforms such as Twitter and Facebook made this process empirically visible, allowing researchers to trace how network structure, homophily, selective exposure, and limited attention shape what users encounter and redistribute \citep{boyd2007social,kwak2010twitter,weng2012competition,wang2020public}. Early computational ranking likewise drew on network structure, as exemplified by PageRank's graph-based ordering of the web \citep{page1999pagerank}. The advent of the artificial intelligence (AI) era has reshaped this network-centric architecture: contemporary social feeds are no longer purely network-driven. They are hybrid systems in which social ties supply candidate information, while AI-driven ranking algorithms reorder, filter, and supplement exposure according to inferred user preferences and predicted engagement \citep{covington2016deep,zhang2019deep,twitter2016timeline}.

In these hybrid systems, platform audits and field experiments show that algorithmic recommendation reshapes the political and topical composition of users' exposure \citep{bakshy2015exposure,huszar2022algorithmic,guess2023feed,gonzalez2023asymmetric,nyhan2023likeminded,haroon2023auditing,hosseinmardi2024causal}. However, these changes do not have a single directional effect on information diversity. Diversity tends to decline when homophilous networks concentrate diffusion among like-minded users and recommendation systems reinforce prior behavior through similarity-based matching or engagement optimization \citep{nguyen2014exploring,schmidt2017anatomy,cinelli2021echo,piao2023human,milli2025engagement}. By contrast, diversity can increase when online intermediaries expand incidental exposure beyond habitual news diets or when recommendation pathways draw users away from ideological extremes \citep{scharkow2020intermediaries,ibrahim2023youtube}. The key question is therefore when algorithmic mediation broadens rather than narrows diversity, especially in hybrid feeds where network propagation, algorithmic ranking, and user sharing behavior jointly shape the information ecology \citep{santos2021link}.

Answering this question requires a mechanistic account, yet existing research omits key parts of the hybrid process. Studies of recommender systems and human--AI feedback show how adaptive recommendation, engagement optimization, and link recommendation can homogenize behavior, amplify bias, and reshape polarization, but generally omit content propagation and resharing through social ties \citep{santos2021link,piao2023human,glickman2025feedback,milli2025engagement}. Empirical studies provide complementary evidence, but typically treat feeds as either chronologically ordered or algorithmically ranked \citep{guess2023feed}, leaving the strength of algorithmic mediation unquantified as a continuous variable \citep{bakshy2015exposure,huszar2022algorithmic,gonzalez2023asymmetric,nyhan2023likeminded,haroon2023auditing,hosseinmardi2024causal}. Diversity-oriented studies incorporate diversity into recommender design or test interventions that broaden exposure and improve information quality, but treat diversity as a target or outcome without an explicit mechanistic model of how it emerges from network propagation, recommendation, and user sharing \citep{stray2021designing,bhadani2022diversity,mattis2024nudging,yu2024nudging,brady2026redesigning}. 

The coupled dynamics of hybrid feeds therefore remain unresolved, and may be nonlinear and non-monotonic. If algorithmic recommendation is too weak, users remain constrained by the topical composition of their local network neighborhoods. If it is too strong, exposure may concentrate toward inferred information diets. Between these extremes, modest algorithmic mediation may complement the social network by surfacing topics that are weakly represented in a user's neighborhood. Moreover, average diversity may hide critical information inequalities. A platform can increase population-level diversity while concentrating that diversity among a subset of users. The equality of topical diversity across users is thus a central outcome.

Here we test this argument using retweeting activity from $18{,}076$ users who were active throughout 2014--2018 and retweeted $549{,}675$ messages from Members of Congress across policy topics. We first measure yearly changes in users' topical sharing around Twitter's 2016 introduction of algorithmic ranking. We then develop a hybrid human--AI diffusion model in which exposure combines network propagation and algorithmic recommendation through a mediation-strength parameter. The model shows that algorithmic mediation has a non-monotonic effect: modest mediation can increase average topical diversity and reduce inequality in diversity across users, whereas stronger mediation narrows sharing and widens disparities. By fitting the model to yearly empirical patterns, we identify the mediation levels under which recommendation broadens rather than contracts the informational ecology of hybrid social feeds.

\section*{Results}
\subsection*{Empirical observations of retweeting behavior on Twitter}
As empirical evidence for the hybrid human--AI system, we first analyzed retweeting behavior on Twitter/X for $18{,}076$ normal users who were active in every year from $2014$ through $2018$.
During this period, these users generated $2{,}290{,}617$ retweets of $549{,}675$ original tweets posted by $694$ Members of Congress.
The original tweets were assigned to $21$ topical categories, defined as $20$ Comparative Agendas Project policy topics plus a residual non-policy category, spanning areas such as health, education, environmental policy, defense, international affairs, and other public issues.
The data source and processing steps are drawn from~\cite{chang2025liberals} and described in Methods.

\begin{figure}[H]
    \centering
    \includegraphics[width=0.5\linewidth]{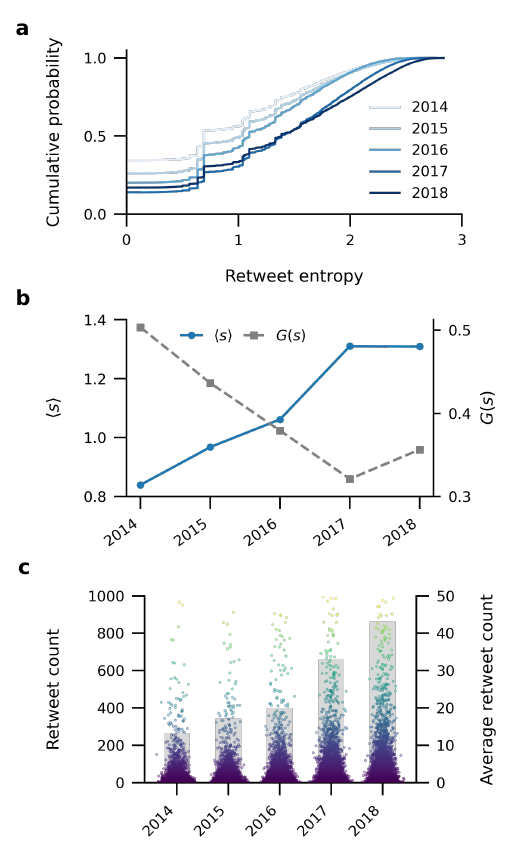}
    \caption{\textbf{Empirical retweet entropy and magnitude on Twitter from 2014 to 2018.}
    \textbf{a}, Empirical cumulative distributions of user retweet entropy over years.
    \textbf{b}, Yearly average retweet entropy, $\langle s\rangle$, and Gini coefficient of retweet entropy, $G(s)$.
    \textbf{c}, Scatters of yearly user retweet counts by year; gray bars of yearly average retweet counts.}
    \label{fig_01}
\end{figure}

To quantify how broadly each user redistributed political information in online social media, we measured user-level retweet entropy from the yearly distribution of a user's retweets across the $21$ topics.
For user $i$, let $f_{i}^{(I)}$ denote the fraction of that user's retweets assigned to topic $I$, and let $K$ be the total number of topics.
The entropy is defined as
$$s_i=-\sum_{I=1}^{K}f_{i}^{(I)}\ln f_{i}^{(I)},$$ 
with larger values indicating retweeting across a broader set of topics.
The empirical cumulative probability distributions of user retweet entropy in Fig.~\ref{fig_01}a shift overall from left to right from $2014$ to $2018$, showing that users increasingly retweeted across more topics.
Notably, curves of $2017$ and $2018$ intersect, which means pronounced inequality in users' retweet entropy emerges during this time period.

Furthermore, we characterize the entropy distribution using its population mean and Gini coefficient.
The average entropy, $\langle s\rangle=\sum_{i=1}^{U}s_i/U$ with $U$ denoting the number of users, measures the typical breadth of topics retweeted by a user.
The Gini coefficient, $G(s)=\sum_{i=1}^{U}\sum_{j=1}^{U}|s_i-s_j| / (2U^2\langle s\rangle)$, quantifies inequality in the distribution of retweet entropy across users.
A larger Gini coefficient indicates a more unequal distribution, where retweet diversity is concentrated among fewer users, while a smaller Gini coefficient represents a more even distribution.
As shown in Fig.~\ref{fig_01}b, the average entropy $\langle s\rangle$ increased from $0.839$ in $2014$ to $1.310$ in $2017$ and then remained nearly unchanged in $2018$ ($1.309$), suggesting that the diversification of retweeted information reached a plateau after $2017$.
In contrast, $G(s)$ decreased from $0.503$ in $2014$ to $0.321$ in $2017$, before rebounding slightly to $0.356$ in $2018$.
Retweet entropy became more evenly distributed across users up to $2017$, but that inequality in users' information diversity re-emerged in $2018$.

Two concurrent changes provide context for these empirical shifts: the introduction of recommendation algorithms on the system side and the increase in engagement on the user side.
First, Twitter introduced an algorithmically ranked timeline in $2016$ \citep{twitter2016timeline}.
Recommendation systems personalize messages to users by analyzing their prior behavior, such as exposure and interaction histories, to rank content that is predicted to be relevant to users \citep{covington2016deep,zhang2019deep}.
Before this change, information diffusion on Twitter resembled an online social network, where users encountered, liked, and retweeted messages posted by accounts they followed chronologically.  After $2016$, information diffusion was jointly shaped by the follower network and the recommendation algorithm, so retweeting reflected both social exposure and algorithmically personalized exposure.
Second, user engagement increased substantially over the same period, particularly as shown in user sharing. 
Average retweet counts rose from $13$ per user in $2014$ to $43$ in $2018$ (Fig.~\ref{fig_01}c). Meanwhile, the share of highly active users increased, whereas that of users with limited retweeting activity declined (Fig.~S2), indicating a shift toward more intensive propagation of political messages.

The empirical change in the retweet entropy distribution is the consequence of the interaction between user behavior and platform architecture.
The social network exposes users to messages retweeted by accounts they follow, while the recommendation algorithm tailors exposure according to inferred user preferences.
User sharing behavior then feeds back into both layers, where retweeted messages become visible to followers through the network, and behavioral signals update the algorithm's estimate of user preferences.
How users, social networks, and recommendation algorithms jointly reshape the information ecosystem on social media therefore requires a deeper mechanistic explanation.

\subsection*{Hybrid human--AI information diffusion model}
\begin{figure}[H]
    \centering
    \includegraphics[width=0.8\linewidth]{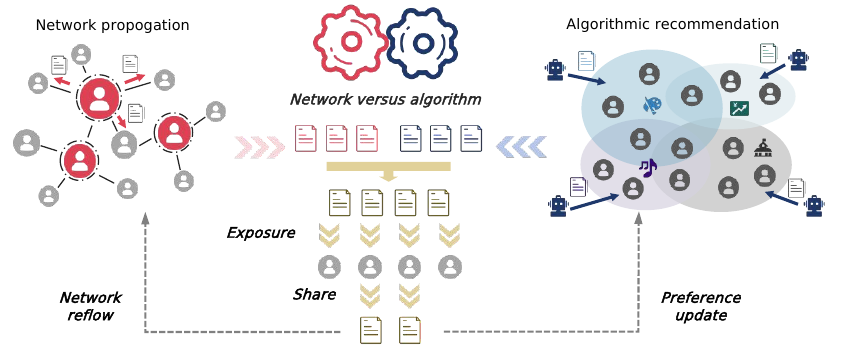}
    \caption{\textbf{Illustration of hybrid human--AI information diffusion model.}
    Information reaches users through network propagation and algorithmic recommendation.
    Users are exposed to messages selected from the hybrid propagation distribution and retweet a subset of them.
    Retweeted messages feed back into the social network and are used to update the estimated user preference.}
    \label{fig_02}
\end{figure}

Motivated by the empirical observations above, we establish a hybrid human--AI model to reproduce the spreading dynamics of information in online social media, grounded in both the social network and the recommendation algorithm (Fig.~\ref{fig_02}).
Because media elites play a pivotal role in producing and framing public issues, while normal users popularize these issues through retweeting, we focus on how elite-generated posts are accessed and spread by normal users.
This setting is consistent with our empirical data, in which normal users retweet messages originally posted by Members of Congress.
Consider an online social network comprising $M$ media elites and $U$ normal users, where normal users may follow one another and may also follow media elites.
When a media elite posts a message or a normal user retweets it, the message can be exposed to the user's followers through the social network.
For simplicity, we formulate the model in the topic level, rather than tracking individual messages.

Assuming that media elites collectively generate messages over $K$ alternative topics, the hybrid human--AI system runs at time $t$ following the steps below.
\begin{enumerate}[label=(\arabic*)]
    \item Network-algorithm hybrid propagation: Network propagation from immediate neighbors, including media elites, has topic distribution $\mathbf{n}_i(t)$ for user $i$, while algorithmic propagation follows the estimated preference distribution $\mathbf{u}_i(t)$ over the $K$ topics. 
    Incorporating both forces, the platform selects messages for each user through the recommendation algorithm with probability $\rho$ and through the social network otherwise.
    Here $\rho\in[0,1]$ describes the algorithmic mediation strength, and thus the topic distribution of hybrid system propagation for each user $i$ is $\mathbf{h}_i(t)=\rho\mathbf{u}_i(t)+(1-\rho)\mathbf{n}_i(t)$.
    \item User sharing: Each user is exposed to $E$ pieces of messages for consumption according to hybrid propagation from the social platform. 
    After that, each user forwards each exposed message with a constant share rate, $\gamma \in (0,1)$, which is typically small.
    \item System update: Let $\mathbf{r}_i(t)$ be the topic distribution of user $i$'s retweets in the current round. 
    User retweets feed back into both layers of the system. 
    First, retweeted messages become available to followers and update the network-propagation state $\mathbf{n}_i(t+1)$ with the time decay rate $\lambda\in(0,1)$. 
    The messages from the latest round are shown to users' windows with the probability $\lambda$ and otherwise from the historic rounds.
    Second, the same retweet distribution updates the estimated user preference by the algorithm according to $\mathbf{u}_i(t+1)=\alpha\mathbf{r}_i(t)+(1-\alpha)\mathbf{u}_i(t)$, where $\alpha\in(0,1)$ is the preference update rate. 
    In particular, if $i$ does not retweet any content in this round, {e.g.}, $\mathbf{r}_i(t)=\mathbf{0}$, we then set $\mathbf{u}_i(t+1) = \mathbf{u}_i(t)$.
    The updated states jointly determine system propagation in the next round.
\end{enumerate}

This procedure is iterated over successive $T$ rounds.
Mathematical descriptions of this model are detailed in Methods and Supplementary Information.
In doing so, our model synthesizes the interplay among the social network, the recommendation algorithm, and the user behavior to offer a mechanistic analysis of information diffusion and diversity in hybrid systems.

\subsection*{Effect of user sharing on information diversity and equality}
We first study how user sharing behavior, including exposure and then sharing, shapes the diversity of information disseminated in online social media.
Hereafter, we quantify information diversity using the normalized entropy, $\tilde{s}=s/\ln K$ with $0\leq\tilde{s}\leq1$, to remove the dependence on the number of topics.
Unless otherwise stated, entropy hereafter refers to this normalized quantity.
We begin with the purely network-driven spread of messages by setting the algorithmic mediation strength to $\rho=0$.
As such, we employ the mean-field approximation that network propagation of information to each user is uniform across the $K$ topics in any time round.
This approximation yields the population distribution of cumulative retweet entropy (see Methods for detailed derivations), which closely matches the simulation outcomes in Fig.~S3.
Subsequently, we estimate the average and Gini coefficient of the cumulative retweet entropy distribution, without algorithmic mediation ($\rho=0$), respectively as
\begin{align}
\langle {\tilde{s}} \rangle
& \simeq
1-\frac{K-1}{2\gamma E T\ln K}, \\
G(\tilde{s}) 
& \simeq
\frac{\sqrt{K-1}}
{\sqrt{2\pi}\left[\gamma E T\ln K-(K-1)/2\right]}.
     \label{eq:engagement-gini-approx}
\end{align}
Here, $\gamma E T$ represents the expected cumulative number of retweets across $T$ time rounds.
Although a longer observation window $T$ can also increase the total number of retweets, users typically engage in social media within a finite time window in reality.
For a fixed window, the analytical results explicitly show a larger sharing number can elevate the average level of information entropy, captured by a higher $\langle \tilde{s} \rangle$, and simultaneously curb inequality of information diversity, reflected by a lower $G(\tilde{s})$.

\begin{figure}[H]
    \centering
    \includegraphics[width=0.5\linewidth]{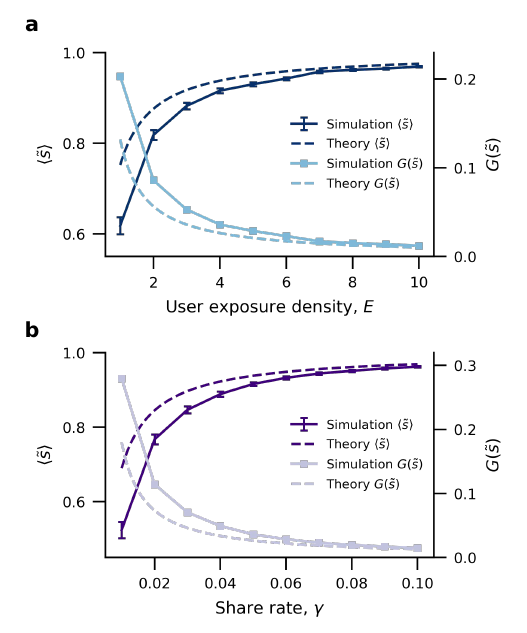}
    \caption{\textbf{Effect of user sharing on cumulative retweet entropy.}
    \textbf{a}, Average cumulative retweet entropy, $\langle\tilde{s}\rangle$ (left axis), and its Gini coefficient, $G(\tilde{s})$ (right axis), as functions of user exposure density $E$.
    \textbf{b}, $\langle\tilde{s}\rangle$ and $G(\tilde{s})$ as functions of share rate $\gamma$.
    Solid and dashed lines denote simulations and analytical predictions, respectively; error bars show $99\%$ confidence intervals over $10$ independent runs.
    Parameters: $K=5$, $M=20$, $U=989$, $\rho=0$, $\alpha=\lambda=0.2$, and $T=100$, with $\gamma=0.05$ in \textbf{a} and $E=4$ in \textbf{b}.}
    \label{fig_03}
\end{figure}

Next, we simulate this process of information propagation over $K=5$ topics throughout a social network consisting of $20$ media elites and $989$ users.
As shown in Fig.~\ref{fig_03}, simulations broadly reproduce the theoretical trends across different scales of user exposure densities and share rates. 
This systematic discrepancy between simulations and analytical predictions still arises because our approximation neglects the heterogeneity in network propagation.
Cross-topic imbalances accumulate over successive rounds of diffusion, generating less uniform exposure and thus lower average entropy, while variation in topic exposure across users broadens the distribution of individual entropy and increases its inequality.
When $E$ and $\gamma$ are small, stronger finite-sampling fluctuations further widen this gap.
As a consequence, our theory tends to overestimate average information entropy and underestimate its Gini coefficient when user sharing magnitude changes. 

Moreover, user sharing behavior, modulated by either $E$ or $\gamma$, affects both the average entropy and its inequality in a nonlinear way.
When the expected number of total sharing remains low, stimulating user sharing sharply raises the average cumulative retweet entropy and dramatically reduces its Gini coefficient (Fig.~\ref{fig_03}).
As $E$ increases from $1$ to $5$ and then to $10$, the simulated average entropy climbs from approximately $0.62$ to $0.93$, before approaching saturation at $0.97$.
In parallel, the Gini coefficient drops steeply from approximately $0.20$ to $0.03$, followed by only a slight further decline to $0.01$.
The share rate shows a similar saturation pattern as well.
Thus, early growth in social media participation can substantially facilitate information diversity and equality, while additional engagement yields smaller gains once exposure and sharing are already high.

This raises a further question: does the same effect of user sharing persist once recommendation algorithms mediate part of the propagation process?
We first examine this question analytically by neglecting cross-user heterogeneity and conditioning on a user's hybrid propagation distribution $\mathbf{h}_i = [h_i^{(I)}]_{I \in [1,K]}$.
Let $S(\cdot)$ denote the entropy of a topic-based probability distribution.
For the empirical retweet topic distribution $\hat{\mathbf{r}}_i$, a second-order approximation gives
\begin{equation}
    \mathbb{E}\!\left[
    S\!\left(\hat{\mathbf{r}}_i\right)
    \mid \mathbf{h}_i
    \right]
    \approx
    S(\mathbf{h}_i)
    -
    \frac{K-1}{2\gamma E},
    \label{eq:main-entropy-bias}
\end{equation}
and
\begin{equation}
\begin{aligned}
    & \mathrm{Var}\!\left[
    s_i
    \mid \mathbf{h}_i
    \right] \\
    & \approx
    \frac{1}{\gamma E}
    \left[
    \sum_{I=1}^{K}
    h_i^{(I)}
    \big(\log h_i^{(I)}\big)^2
    -
    \bigg(
    \sum_{I=1}^{K}
    h_i^{(I)}
    \log h_i^{(I)}
    \bigg)^2
    \right].
\end{aligned}
    \label{eq:main-entropy-variance-explicit}
\end{equation}
The detailed derivations are provided in Supplementary Information.
Because this approximation neglects cross-user heterogeneity, the statistics above for a specific user can be used to estimate the population-level trend.
These expressions show that, for a fixed hybrid propagation distribution, increasing sharing raises expected retweet entropy by reducing finite-sampling bias and narrows the entropy distribution by reducing sampling variance.

Algorithmic mediation, however, changes the hybrid propagation distribution itself and therefore progressively weakens the nonlinear benefit of early growth in user sharing.
To test this interaction, we repeated the analysis in an empirically grounded setting using observed posts over $21$ topics generated by $694$ media elites and propagated by $18{,}076$ users from 2014 to 2018.
When algorithmic mediation is below the algorithm-dominated regime, increasing total sharing, $E\gamma T$, still produces a higher and narrower entropy distribution across users, but the sharp low-sharing gain becomes less pronounced as $\rho$ increases (Fig.~S4).
When mediation becomes extremely strong, this relationship can reverse: for $\rho>0.96$, greater sharing reduces information diversity and increases inequality across users (Fig.~S5).
This reversal occurs because additional sharing then samples an increasingly concentrated preference-driven distribution, making some users' retweets more homogeneous rather than more diverse.

\begin{figure}[H]
    \centering
    \includegraphics[width=0.95\linewidth]{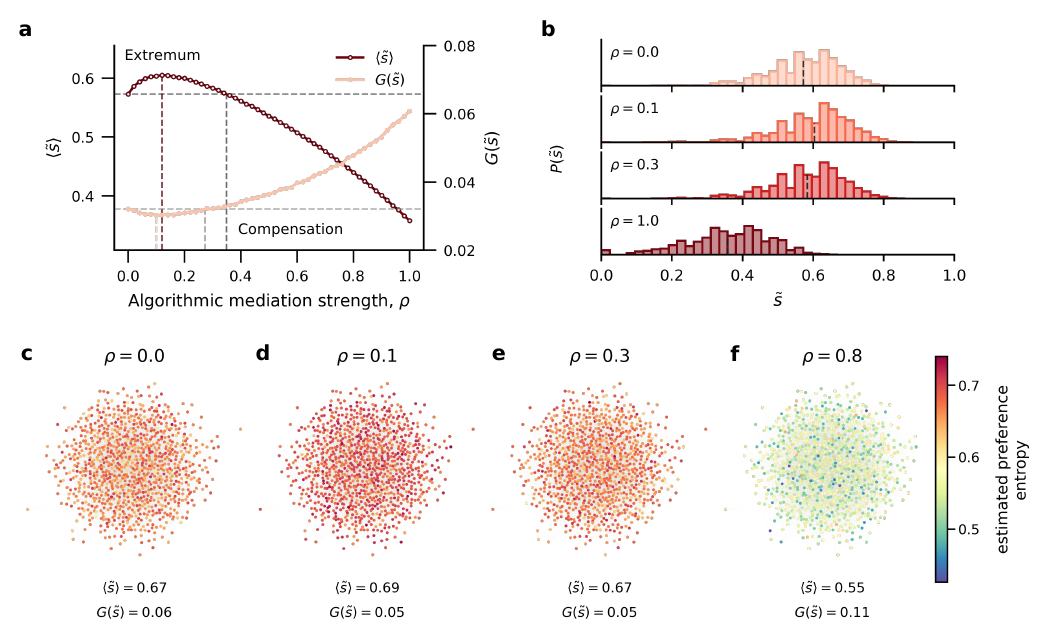}
    \caption{\textbf{Effect of algorithmic mediation strength on information entropy.}
    \textbf{a}, Average cumulative retweet entropy, $\langle\tilde{s}\rangle$, and its Gini coefficient, $G(\tilde{s})$, as functions of algorithmic mediation strength $\rho$.
    Vertical red dashed lines indicate their extrema, and horizontal dashed lines mark the corresponding $\rho=0$ baselines used to define compensation points.
    Results are averaged over $10$ independent simulations.
    \textbf{b}, Distributions of cumulative retweet entropy at selected values of $\rho$.
    Dashed lines indicate the means.
    \textbf{c}--\textbf{f}, Sampled user--user networks at $\rho=0$, $0.1$, $0.3$, and $0.8$, with nodes colored by estimated preference entropy.
    Simulations use the aggregated empirical posts from 2014 to 2018 with $K=21$, $M=694$, and $U=18{,}076$.
    Parameters: $\lambda=0.2$, $\alpha=0.2$, $E=10$, $\gamma=0.01$, and $T=100$.}
    \label{fig_04}
\end{figure}

We next vary the two engagement parameters, the exposure density $E$ and the share rate $\gamma$. Raising $E$ increases $\langle s\rangle$ and decreases $G(s)$, both saturating by roughly $E=6$ (Fig.~\ref{fig_04}a,b): users exposed to more messages per round sample more topics, which broadens and equalizes sharing. This effect interacts with $\rho$. Under network-dominated diffusion (small $\rho$) more exposure continues to raise diversity, whereas under algorithm-dominated diffusion (large $\rho$) the gain from additional exposure is muted, because ranking re-concentrates attention on already-preferred topics regardless of volume (Fig.~\ref{fig_04}c,d). The joint $(E,\gamma)$ maps in Fig.~\ref{fig_04}e,f show that the share rate amplifies the same pattern: higher $\gamma$ accelerates preference reinforcement, so its effect on $\langle s\rangle$ and $G(s)$ runs in the same direction as $E$ but depends on $\rho$.

\subsection*{Effect of algorithmic mediation on information diversity and equality}
To study how algorithmic mediation shapes information diversity and equality in online social media, we first conducted simulations using empirical posts from 2014 to 2018 generated by $694$ media elites and spread by $18{,}076$ normal users.
The elite--user layer is represented as a bipartite network with connection probability $0.05$, and the user--user layer is represented as an Erd\H{o}s--R\'enyi network with edge probability $0.01$ and mean degree $180$.
The simulations reveal a non-monotonic effect of algorithmic mediation on average cumulative retweet entropy (Fig.~\ref{fig_04}a).
Starting from $\langle\tilde{s}\rangle\approx0.57$ in the absence of algorithmic mediation ($\rho=0$), the average entropy rises to a maximum of $0.61$ at $\rho \approx 0.12$, before declining steadily to $0.36$ as $\rho$ reaches $1$.
This optimum identifies the level of algorithmic mediation that best promotes information diversity under the current sharing level.
We also identify a compensation point at $\rho\approx0.35$, where the average entropy returns to its network-only baseline.
Above this compensation point, algorithmic mediation produces less diverse information ecology than no algorithmic intervention, thus giving rise to the formation of information cocoons.

The Gini coefficient follows the same logic yet with the direction reversed.
Information equality is optimized at $\rho \approx 0.10$, where $G(\tilde{s})$ is minimized.
At this point, the retweet entropy distribution is relatively concentrated within an intermediate-to-high range approximately from $0.4$ to $0.8$ (Fig.~\ref{fig_04}b), indicating that many users retweet across a relatively diverse set of topics.
In contrast, when algorithmic mediation dominates at $\rho=1$, the distribution shifts left and spreads over a lower range spanning from $0$ to $0.6$, indicating that some users become much more topic-homogeneous than others.
The compensation point for inequality occurs at $\rho\approx0.27$.
Beyond this point, algorithmic mediation makes retweet entropy more unequal than in the network-only baseline.

This phenomenon reflects a tension between network propagation and algorithmic propagation.
For user $i$, the hybrid propagation distribution can be written as
$\mathbf{h}_i(\rho)=\rho\mathbf{u}_i+(1-\rho)\mathbf{n}_i$, where $\mathbf{u}_i$ is the preference-driven distribution and $\mathbf{n}_i$ is the network-driven distribution.
The local effect of increasing algorithmic mediation is governed by
\begin{equation}
        \frac{\partial S(\mathbf{h}_i)}{\partial \rho}
    =
    -
    \sum_{I=1}^{K}
    \left(u_i^{(I)}-n_i^{(I)}\right)\log h_i^{(I)}.
\end{equation}
This derivative shows that the effect of $\rho$ depends on the current position of $\mathbf{h}_i(\rho)$ along the path from network-driven exposure $\mathbf{n}_i$ to preference-driven exposure $\mathbf{u}_i$.
When $\rho$ is small, adding a modest algorithmic component can complement network propagation by exposing users to preference-relevant topics that are weakly represented in their social neighborhoods, globally increasing information diversity and reducing cross-user discrepancy.
When $\rho$ is large, however, the hybrid distribution becomes increasingly aligned with user preferences, which are more concentrated and heterogeneous (Fig.~\ref{fig_04}f); further mediation then narrows individual retweet distributions and increases inequality.
Because the entropy of this mixture is concave in $\rho$,
\begin{equation}
        \frac{\partial^2 S(\mathbf{h}_i)}{\partial \rho^2}
    =
    -
    \sum_{I=1}^{K}
    \frac{(u_i^{(I)}-n_i^{(I)})^2}{h_i^{(I)}}
    \leq 0,
\end{equation}
these opposing forces naturally produce an interior optimum and a later compensation point.


\subsection*{Optimal algorithmic mediation for information diversity and equality}
Platforms can tune the strength of algorithmic ranking, but the resulting level of user engagement ultimately depends on users’ willingness to participate by sharing content.
This raises a practical question: for a given total sharing level, what degree of algorithmic mediation best supports information diversity and equality, and beyond what point does excessive mediation suppress both relative to the network-only baseline?
We therefore repeated the empirical-scale simulations across varying total sharing, $E\gamma T$, and identified two landmarks for each sharing level: the extremum $\rho$ that maximizes $\langle\tilde{s}\rangle$ or minimizes $G(\tilde{s})$, and the compensation $\rho$ at which the corresponding statistic returns to its pure network baseline.

\begin{figure}[H]
    \centering
    \includegraphics[width=0.5\linewidth]{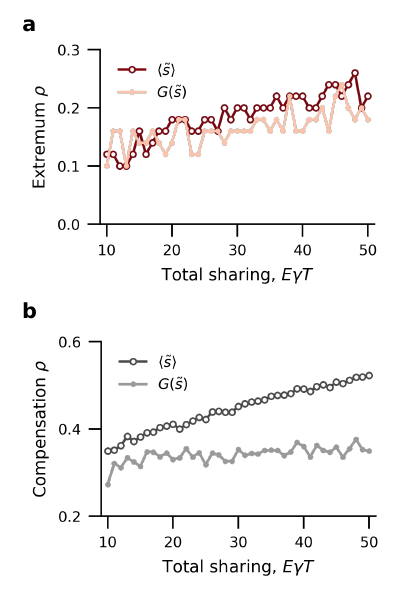}
    \caption{\textbf{Extremum and compensation of algorithmic mediation strength.}
    \textbf{a}, Values of $\rho$ that maximize mean cumulative retweet entropy, $\langle\tilde{s}\rangle$, or minimize its Gini coefficient, $G(\tilde{s})$, across total sharing $E\gamma T$.
    \textbf{b}, Compensation values of $\rho$ at which the corresponding statistic returns to its $\rho=0$ baseline.
    Simulations follow the empirical setting and parameters of Fig.~\ref{fig_04}, with $E\gamma T$ varied from $10$ to $50$.}
    \label{fig_05}
\end{figure}

Fig.~\ref{fig_05}a shows that the optimal algorithmic mediation strengths for both diversity and equality increase with total sharing but remain in a relatively low range, roughly $\rho=0.10$--$0.25$.
This upward shift suggests that greater sharing buffers the concentrating effect of algorithmic mediation, allowing the system to tolerate somewhat stronger mediation before diversity or equality starts to decline.
The compensation points separate more clearly, as shown in Fig.~\ref{fig_05}b.
For average entropy, the compensation $\rho$ rises approximately from $0.35$ to $0.52$ as total sharing grows, again indicating that higher sharing delays the point at which mediation becomes worse than the pure network baseline.
For the Gini coefficient, however, the compensation value is less responsive to variations in user sharing and remains close to $\rho \approx 0.35$.
Overall, increasing engagement broadens the safe range of algorithmic mediation for average diversity, while equality is governed by a more stable and restrictive threshold.

\subsection*{Model-data integration of information diffusion in Twitter}
Finally, we integrate the theoretical model with the empirical Twitter data to examine whether the same mechanism can account for the observed yearly changes in retweet entropy distributions.
This integration is not intended to perfectly reproduce the full empirical trajectory, which is shaped by many platform- and event-specific factors outside the model.
It instead asks whether the model can mechanistically simulate and explain the relative movement of average entropy and entropy inequality over time.
To reduce systematic level differences between empirical and simulated entropy, we normalize both series by their 2014 values.
For each year from 2014 to 2018, we use the corresponding empirical elite posts as model input and fit the algorithmic mediation strength $\rho_t$ and the share rate $\gamma_t$ to match the observed trends in $\langle s\rangle$ and $G(s)$ as closely as possible.
The data fitting algorithm is detailed in Methods.

\begin{figure}[H]
    \centering
    \includegraphics[width=\linewidth]{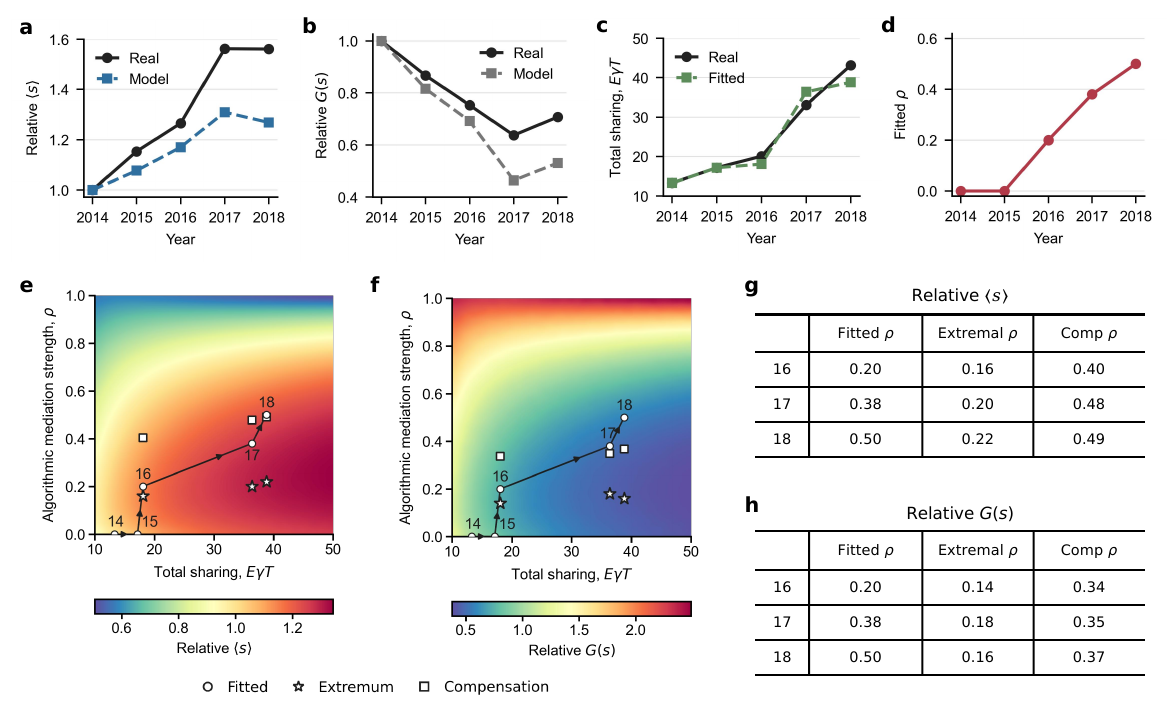}
    \caption{\textbf{Model-data integration of retweet entropy and algorithmic mediation from 2014 to 2018.}
    \textbf{a}--\textbf{b}, Empirical and simulated average retweet entropy, $\langle s\rangle$ (\textbf{a}), and its Gini coefficient, $G(s)$ (\textbf{b}), relative to 2014. Model simulations are averaged over $10$ runs.
    \textbf{c}--\textbf{d}, Empirical and fitted total sharing, $E\gamma T$ (\textbf{c}), and fitted algorithmic mediation strength, $\rho$ (\textbf{d}).
    \textbf{e}--\textbf{f}, Relative model-predicted $\langle s\rangle$ (\textbf{e}) and $G(s)$ (\textbf{f}) across total sharing and $\rho$. Circles show fitted yearly positions, stars correspond to extrema, and squares represent compensation points at which values return to their $\rho=0$ baselines.
    \textbf{g}--\textbf{h}, Fitted, extremal, and compensation $\rho$ for $\langle s\rangle$ (\textbf{g}) and $G(s)$ (\textbf{h}).
    The calibration uses the empirical posts for the corresponding year, and heatmap simulations use the aggregated empirical posts from 2014 to 2018.
    Parameters: $K=21$, $M=694$, $U=18{,}076$, $\lambda=0.2$, $\alpha=0.2$, $E=10$, and $T=100$.}
    \label{fig_06}
\end{figure}

Under the yearly best-fitted configurations, the model reproduces the main temporal patterns in the empirical data (Fig.~\ref{fig_06}a--b).
Average entropy increases until 2017 and then levels off, while the Gini coefficient declines through 2017 before rising again in 2018.
The simulated trajectories remain lower than the empirical trajectories, partly due to real-world entropy heterogeneity generated by additional sources not included in the theoretical model, such as individual variation in users' share rates.
The fitted total sharing closely tracks the observed average retweet count per user (Fig.~\ref{fig_06}c).
The fitted algorithmic mediation strength is zero before the introduction period and then rises steadily after 2016, reaching $\rho=0.50$ by 2018 (Fig.~\ref{fig_06}d).

To interpret these fitted trajectories mechanistically, we place the fitted yearly positions on model-predicted landscapes generated from the aggregated elite posts from 2014 to 2018 (Fig.~\ref{fig_06}e--f).
We then mark the model-predicted extrema as well as the compensation points across different scales of user sharing (Fig.~\ref{fig_06}e--h).
Regarding average entropy, the fitted positions fall between the diversity optimum and the compensation boundary, with 2018 approaching that boundary; in this scenario, algorithmic mediation can still increase information diversity relative to the network-only baseline.
Regarding the Gini coefficient, however, the fitted positions in 2017 and 2018 move beyond the compensation boundary, indicating that this mediation intensity would cause inequality in users' information diversity.
These comparisons jointly suggest that mediation strengths around $\rho=0.15$--$0.20$ can improve diversity while preserving equality, offering a model-based guideline for designing healthier information ecosystems.

\section*{Discussion}

We modeled information diffusion on social platforms as a process on two
coupled layers: social network and algorithmically ranked recommendation, and measured its consequences for the diversity of sharing.
Network propagation alone produces broad and relatively equal sharing.
Algorithmic recommendation reshapes it non-monotonically. A modest algorithmic
component complements the social network by surfacing preference-relevant
topics that are weakly represented in a user's neighborhood, raising both
average diversity and its equality across users. 
Beyond a compensation point, ranking increasingly samples users' own concentrated preferences, narrowing individual sharing and widening differences between users.

Throughout over two millions of retweets of U.S. Members of Congress, the diversity of user
sharing rose after 2016 while its inequality across users rose alongside it.
The model reproduces this joint pattern when an increasing share of exposure
is attributed to algorithmic rather than network mediation, with the fitted
share reaching approximately $0.50$ by 2018. This non-monotonicity is also
consistent with mixed evidence on filter bubbles. Audit studies find that
algorithmic ranking adds only a modest reduction in cross-cutting exposure
beyond users' self-selection \citep{bakshy2015exposure}, whereas popular
accounts predict severe narrowing \citep{pariser2011filter}. In our framework
both observations are compatible, because the effect of mediation depends on
its strength relative to user engagement, and moderate mediation can exceed
the network-only baseline in diversity. Our model--data integration suggests
that Twitter crossed the equality compensation point around 2017 while
remaining within the diversity-enhancing range, so that average diversity rose
even as users diverged into high- and low-diversity groups, the bifurcation
visible in Fig.~\ref{fig_01}a. Read through the supply-and-demand framing of
attention \citep{chang2025liberals}, the recommender expands the supply of
reachable topics while concentrating realized demand.

The results also have implications for platform design. This distributional
perspective matters because algorithmic systems can reproduce or intensify
existing social inequalities \citep{bui2025algorithmic}. Because the diversity
optimum and the equality compensation point occur at low mediation strengths
($\rho \approx 0.10$--$0.35$) and increase only slowly with engagement,
mediation near $\rho=0.15$--$0.20$ improves diversity while preserving
equality. This guideline is model-based and conditional on the fitted
engagement regime rather than a causal policy estimate, but it suggests that
diversity-aware design requires tempering recommendation rather than
abandoning it.

Several limitations bound these conclusions. First, identification rests on
the timing of algorithmic ranking. Twitter introduced its ranked timeline in
2016 \citep{twitter2016timeline}, but 2016 also brought a presidential
election and a sharp rise in retweet volume (Fig.~\ref{fig_01}c), and our
design cannot separate these co-occurring changes. The fitted $\rho$ should
therefore be read as the ranking share consistent with the data under the
model. Second, we measure topical diversity of congressional content rather
than users' full information diets, so low entropy indicates concentrated
retweeting of elite posts rather than an information cocoon per se. This focus
complements evidence that political virality also varies across identity,
policy, and affective dimensions \citep{changfang2024taiwan}. Third, the
model assumes homogeneous engagement parameters and a static, topically
non-homophilous network. Although social ties are ideologically homophilous,
they need not be topically narrow, since copartisans discuss a wide range of
issues; topically homophilous networks would likely shift the optimum and
remain a question for future work. Finally, we exploit only the incidence
structure of the user--topic hypergraph. Further integration of the hypergraph
framework, including higher-order contagion across overlapping hyperedges
\citep{benson2016higher,iacopini2019simplicial,battiston2020networks} and
co-evolution of the network and topic layers as the recommender suggests ties
as well as content, may deepen the understanding of these dynamics.

\section*{Materials and Methods}

\subsection*{Data Collection} 
{We build on the corpus and validated annotations assembled in prior work \citep{chang2025liberals}, itself seeded by the Member-of-Congress (MC) tweet collection of \citep{frimer2023incivility}.
We tracked 18{,}076 users who were active in each year from 2014 through 2018. These users generated 2{,}290{,}617 retweets of 549{,}675 distinct tweets posted by 694 Members of Congress (MC). We then assigned each original MC tweet to a policy topic.
Topic labels were obtained with a BERTweet classifier trained on Comparative Agendas Project (CAP) codes, which maps each tweet to one of the 20 CAP policy categories or to a residual non-policy/other category.
We therefore use a fixed topic set with $K=21$ categories throughout the analysis: non-policy/other, macroeconomics, civil rights, health, agriculture, labor and employment, education, environment, energy, immigration, transportation, law and crime, social welfare, community development and housing, banking and finance, defense, science and technology, foreign trade, international affairs, government operations, and public lands and water management.
Each retweet inherits the topic label of the original MC tweet that it retweets, allowing us to construct user-level retweet distributions over topics.
For each user and year, we aggregated retweets by topic to obtain the empirical retweet-topic distribution, from which we computed retweet entropy, mean entropy, and the Gini coefficient of entropy across users.}

\subsection*{Model descriptions} 
We model the information propagation, driven by both the social network and the recommendation algorithm, over $U$ normal users and $K$ topics.
For user $i$ at round $t$, let $\mathbf{u}_i(t)$, $\mathbf{n}_i(t)$, $\mathbf{h}_i(t)$ and $\hat{\mathbf{r}}_i(t)$ denote the estimated user preference, network propagation, hybrid propagation and observed retweet distributions over topics, respectively.
The platform combines network-driven and algorithm-driven propagation as
\begin{equation}
        \mathbf{h}_i(t)=\rho\mathbf{u}_i(t)+(1-\rho)\mathbf{n}_i(t),
\end{equation}
where $\rho \in [0,1]$ is the algorithmic mediation strength.
The network propagation and estimated preference are updated from the observed retweet distribution by
\begin{align}
    \mathbf{n}_i(t+1)
    & =
    \lambda\sum_{j=1}^{U}P_{ij}\hat{\mathbf{r}}_j(t)
    +(1-\lambda)\mathbf{n}_i(t), \\
    \mathbf{u}_i(t+1)
    & =
    \alpha\hat{\mathbf{r}}_i(t)
    +(1-\alpha)\mathbf{u}_i(t).
\end{align}
where $P_{ij}$ weighs how much user $j$'s retweets contribute to user $i$'s network propagation, $\lambda$ is the time decay rate of network propagation, and $\alpha$ is the preference update rate.

In the time round $t$, user $i$ is exposed to $E$ messages sampled from $\mathbf{h}_i(t)$, and each exposed message is independently retweeted with share rate $\gamma$.
Let $L_i(t)$ be the total number of retweets made by user $i$ in that time round, and $\mathbf{Y}_i(t)$ be the corresponding vector of retweet counts by topic.
Conditioned on $L_i(t)>0$, the observed retweet distribution is $\hat{\mathbf{r}}_i(t)=\mathbf{Y}_i(t)/L_i(t)$.
Because the share rate is topic-independent, this retweet sampling does not systematically change topic proportions.
For large effective sample size $\gamma E$, we have
\begin{align}
\hat{\mathbf{r}}_i(t)=\mathbf{h}_i(t)+\boldsymbol{\xi}_i(t),
\end{align}
where $\boldsymbol{\xi}_i(t)$ is the noise introduced by the sampling with the corresponding expectation and covariance as
\begin{equation*}
    \mathbb{E}[\boldsymbol{\xi}_i(t)\mid \mathbf{h}_i(t)]=\mathbf{0},
\end{equation*}
\begin{equation*}
    \mathrm{Cov} \left[\boldsymbol{\xi}_i(t)\mid \mathbf{h}_i(t)\right]
    \approx
    \frac{1}{\gamma E}
    \left[
    \mathrm{diag} \big(\mathbf{h}_i(t)\big)
    -
    \mathbf{h}_i(t)\mathbf{h}_i(t)^\top
    \right].
\end{equation*}
This approximation is conditioned on nonzero retweet counts and thus treats fluctuations in $L_i(t)$ as higher-order error.

\subsection*{Mean-field approximation for cumulative retweet entropy distribution} 
Here, we set the algorithmic mediation strength as $\rho=0$, and hybrid propagation thus reduces to network propagation.
In this purely network-driven limit, we adopt the mean-field approximation with the assumption that the network propagation distribution to each user is completely uniform across topics in each time round, e.g., $\mathbf{n}_i(t)=(1/K,\ldots,1/K)$.
We then pool exposure and retweet sampling over the $T$ rounds at the cumulative level.
Conditioned on a nonzero cumulative retweet count, and replacing the random cumulative count by its expectation $\gamma E T$, the cumulative topic-count vector is approximated as
\[
    \mathbf{C}_i
    =
    \left(C_i^{(1)},\ldots,C_i^{(K)}\right)
    \sim
    \operatorname{Multinomial}
    \left(\gamma E T;\frac{1}{K},\ldots,\frac{1}{K}\right),
\]
with the covariance
\[
    \operatorname{Cov} \left[\mathbf{C}_i\right]
    =
    \gamma E T
    \left[
    \frac{1}{K}\mathbf{I}_K
    -
    \frac{1}{K^2}\mathbf{1}\mathbf{1}^{\top}
    \right],
\]
where $\mathbf{I}_K$ is the identity matrix with dimension $K$ and $\mathbf{1}$ is the $K$-dimensional column vector of ones.
Let $f_i^{(I)}=C_i^{(I)}/(\gamma E T)$ be the cumulative retweet frequency of topic $I$.
We then define the normalized fluctuation $\boldsymbol{\delta}_i$ with each element $\delta_i^{(I)}=f_i^{(I)}-1/K$, whose covariance is
\[
    \operatorname{Cov}\left[\boldsymbol{\delta}_i\right]
    =
    \frac{1}{\gamma E T}
    \left[
    \frac{1}{K}\mathbf{I}_K
    -
    \frac{1}{K^2}\mathbf{1}\mathbf{1}^{\top}
    \right].
\]

Define the normalized cumulative retweet entropy as $\tilde{s}_i=-\sum_{I=1}^{K}f_i^{(I)}\ln f_i^{(I)}/\ln K$, which can then be expanded around the uniform distribution.
Substituting $f_i^{(I)}=1/K+\delta_i^{(I)}$ into this entropy function, the linear term vanishes because $\sum_I\delta_i^{(I)}=0$, and the second-order expansion gives
\[
    \tilde{s}_i
    \approx
    1
    -
    \frac{K}{2\ln K}
    \sum_{I=1}^{K}\left(\delta_i^{(I)}\right)^2.
\]
Under the large-$\gamma E T$ and uniform topic distribution assumptions stated above, this expansion gives the Gaussian estimate for each user $i$:
\begin{equation}
    \tilde{s}_i
    \sim
    \mathcal{N}\!\left(
    \mu_{\tilde{s}_i},
    \sigma_{\tilde{s}_i}^2
    \right),
\end{equation}
with its mean and variance respectively specified as
\begin{align}
    \mu_{\tilde{s}_i}
    & \simeq
    1-\frac{K-1}{2\gamma E T\ln K}, \\
    \sigma_{\tilde{s}_i}^2
    & \simeq
    \frac{K-1}{2(\gamma E T)^2(\ln K)^2}.
\end{align}
Because we neglect the user-level heterogeneity in this approximation, this Gaussian distribution for each user can be also used to approximate the population distribution of cumulative retweet entropy.

Based on the above, the Gini coefficient of cumulative retweet entropy is therefore estimated as
\begin{equation}
    G(\tilde{s})
    \simeq
    \frac{\sigma_{\tilde{s}}}
    {\sqrt{\pi}\,\mu_{\tilde{s}}}.
\end{equation}
Substituting the mean and variance into it gives the specific form as \eqref{eq:engagement-gini-approx}.

\subsection*{Data fitting algorithm}
To reduce the computational complexity of this fitting procedure, we fix the time decay rate of network propagation as $\lambda=0.2$ and the preference update rate as $\alpha=0.2$ throughout the specified period.
Based on these considerations, we seek an optimal yearly configuration of the user total sharing and the algorithmic mediation strength to closely reproduce the evolution of information diversity from 2014 to 2018.
For simplicity, we change the total sharing by tuning the share rate and fixing the exposure density $E=10$ and time round $T=100$.
The empirical Twitter data provides the observed user average total sharing ${r}_t$ for each year $t$ (Fig.~\ref{fig_01}c), from which we derive the empirical sharing rate as $\gamma^{\mathrm{real}}_t={r}_t/(ET)$.
For each single year $t$, we vary the share rate $\gamma_t$ within the vicinity of its empirical range and the algorithmic mediation strength $\rho_t$ from 0 to 1, and then compare the simulated and observed mean and Gini coefficient of retweet entropy.
By setting the statistics from 2014 as the baseline, the loss function of each year $t$ is then defined as 
\begin{equation}
    \mathcal{L}_t
    =
    \left(
    \frac{\langle s\rangle_t^{\mathrm{simu}}}{\langle s\rangle_{2014}^{\mathrm{simu}}}
    -
    \frac{\langle s\rangle_t^{\mathrm{real}}}{\langle s\rangle_{2014}^{\mathrm{real}}}
    \right)^2
    +
    \left(
    \frac{G_t^{\mathrm{simu}}}{G_{2014}^{\mathrm{simu}}}
    -
    \frac{G_t^{\mathrm{real}}}{G_{2014}^{\mathrm{real}}}
    \right)^2,
\end{equation}
where the simulation outcome for each parameter configuration is averaged over 10 realizations.
In particular, $\mathcal{L}_{2014}=0$.

We develop a staged least-squares calibration algorithm to fit the model to the empirical yearly entropy statistics as follows:
\begin{enumerate}[label=(\arabic*)]
    \item \textit{Initialization.} For each year $t=2014,\ldots,2018$, we observe the empirical elite-post topic distribution $\mathbf{POST}_t$, the average retweet entropy, the Gini coefficient of retweet entropy and the user retweet counts from empirical data.
    We construct the elite-to-user network and the user-to-user network. 
    The estimated user preference state is initialized by a Dirichlet preference matrix. 
    
    \item \textit{Fitting the yearly algorithmic mediation strength sequence $\{\rho_t\}$.} 
    We first set $\gamma_t=\gamma_t^{\mathrm{real}}$ for each year $t$. 
    We fix $\rho_{2014}=\rho_{2015}=0$ to represent the pre-recommendation baseline and sweep $\rho_t$ for subsequent years to identify the yearly algorithmic mediation strength that minimizes $\mathcal{L}_t$.
    
    \item \textit{Fitting the yearly share rate sequence $\{ \gamma_t\}$.} 
    Holding $\rho_t$ obtained from the first stage fixed, we then vary $\gamma_t$ around $\gamma_t^{\mathrm{real}}$ to simulate entropy statistics for each year excluding 2014.

    \item \textit{Local refinement.} We jointly search local neighborhoods around $\rho_t$ obtained from the first stage and $\gamma_t$ obtained from the second stage. 
    The fitted trajectory is the set of yearly parameter pairs $(\rho_t,\gamma_t)$ that minimizes the total loss $\mathcal{L}=\sum_{t=2015}^{2018}\mathcal{L}_t$.
\end{enumerate}

\section*{Competing Interests}
The authors declare no competing interests.

\clearpage

\clearpage

\clearpage

\clearpage

\clearpage

\clearpage

\clearpage
\begingroup
\bibliographystyle{naturemag}
\bibliography{pnas_refs}

@book{rogers2003diffusion,
  title={Diffusion of Innovations},
  author={Rogers, Everett M.},
  year={2003},
  publisher={Free Press}
}

@article{granovetter1973strength,
  title={The strength of weak ties},
  author={Granovetter, Mark S.},
  journal={American Journal of Sociology},
  volume={78},
  number={6},
  pages={1360--1380},
  year={1973}
}

@inproceedings{bakshy2012role,
  title={The role of social networks in information diffusion},
  author={Bakshy, Eytan and Rosenn, Itamar and Marlow, Cameron and Adamic, Lada A.},
  booktitle={Proceedings of the 21st International Conference on World Wide Web},
  pages={519--528},
  year={2012}
}

@article{chang2025liberals,
  title={Liberals and conservatives share information differently on social media},
  author={Chang, Ho-Chun Herbert and Druckman, James N. and Ferrara, Emilio and Willer, Robb},
  journal={PNAS Nexus},
  volume={4},
  number={7},
  pages={pgaf206},
  year={2025},
  doi={10.1093/pnasnexus/pgaf206}
}

@article{frimer2023incivility,
  title={Incivility is rising among American politicians on Twitter},
  author={Frimer, Jeremy A. and others},
  journal={Social Psychological and Personality Science},
  volume={14},
  number={2},
  pages={259--269},
  year={2023}
}

@article{vosoughi2018spread,
  title={The spread of true and false news online},
  author={Vosoughi, Soroush and Roy, Deb and Aral, Sinan},
  journal={Science},
  volume={359},
  number={6380},
  pages={1146--1151},
  year={2018}
}

@inproceedings{covington2016deep,
  title={Deep neural networks for YouTube recommendations},
  author={Covington, Paul and Adams, Jay and Sargin, Emre},
  booktitle={Proceedings of the 10th ACM Conference on Recommender Systems},
  pages={191--198},
  year={2016}
}

@misc{twitter2016timeline,
  title={Never miss important Tweets from people you follow},
  author={{Twitter}},
  year={2016},
  month={February},
  day={10},
  howpublished={\url{https://blog.x.com/en_us/a/2016/never-miss-important-tweets-from-people-you-follow}},
  note={Accessed 9 July 2026}
}

@article{weng2012competition,
  title={Competition among memes in a world with limited attention},
  author={Weng, Lilian and Menczer, Filippo and Ahn, Yong-Yeol},
  journal={Scientific Reports},
  volume={2},
  number={1},
  pages={335},
  year={2012}
}

@article{boyd2007social,
  title={Social network sites: Definition, history, and scholarship},
  author={boyd, danah m. and Ellison, Nicole B.},
  journal={Journal of Computer-Mediated Communication},
  volume={13},
  number={1},
  pages={210--230},
  year={2007}
}

@inproceedings{kwak2010twitter,
  title={What is Twitter, a social network or a news media?},
  author={Kwak, Haewoon and Lee, Changhyun and Park, Hosung and Moon, Sue},
  booktitle={Proceedings of the 19th International Conference on World Wide Web},
  pages={591--600},
  year={2010}
}

@article{zhang2019deep,
  title={Deep learning based recommender system: A survey and new perspectives},
  author={Zhang, Shuai and Yao, Lina and Sun, Aixin and Tay, Yi},
  journal={ACM Computing Surveys},
  volume={52},
  number={1},
  pages={1--38},
  year={2019}
}

@article{bakshy2015exposure,
  title={Exposure to ideologically diverse news and opinion on Facebook},
  author={Bakshy, Eytan and Messing, Solomon and Adamic, Lada A.},
  journal={Science},
  volume={348},
  number={6239},
  pages={1130--1132},
  year={2015}
}

@book{pariser2011filter,
  title={The Filter Bubble: What the Internet Is Hiding from You},
  author={Pariser, Eli},
  year={2011},
  publisher={Penguin}
}

@article{huszar2022algorithmic,
  title={Algorithmic amplification of politics on Twitter},
  author={Husz{\'a}r, Ferenc and Ktena, Sofia Ira and O'Brien, Conor and Belli, Luca and Schlaikjer, Andrew and Hardt, Moritz},
  journal={Proceedings of the National Academy of Sciences},
  volume={119},
  number={1},
  pages={e2025334119},
  year={2022},
  doi={10.1073/pnas.2025334119}
}

@article{guess2023feed,
  title={How do social media feed algorithms affect attitudes and behavior in an election campaign?},
  author={Guess, Andrew M. and Malhotra, Neil and Pan, Jennifer and Barber{\'a}, Pablo and Allcott, Hunt and Brown, Taylor and Crespo-Tenorio, Adriana and others},
  journal={Science},
  volume={381},
  number={6656},
  pages={398--404},
  year={2023},
  doi={10.1126/science.abp9364}
}

@article{gonzalez2023asymmetric,
  title={Asymmetric ideological segregation in exposure to political news on Facebook},
  author={Gonz{\'a}lez-Bail{\'o}n, Sandra and Lazer, David and Barber{\'a}, Pablo and Zhang, Meiqing and Allcott, Hunt and Brown, Taylor and Crespo-Tenorio, Adriana and others},
  journal={Science},
  volume={381},
  number={6656},
  pages={392--398},
  year={2023},
  doi={10.1126/science.ade7138}
}

@article{nyhan2023likeminded,
  title={Like-minded sources on Facebook are prevalent but not polarizing},
  author={Nyhan, Brendan and Settle, Jaime and Thorson, Emily and Wojcieszak, Magdalena and Barber{\'a}, Pablo and Chen, Annie Y. and Allcott, Hunt and others},
  journal={Nature},
  volume={620},
  pages={137--144},
  year={2023},
  doi={10.1038/s41586-023-06297-w}
}

@article{yu2024nudging,
  title={Nudging recommendation algorithms increases news consumption and diversity on YouTube},
  author={Yu, Xudong and Haroon, Muhammad and Menchen-Trevino, Ericka and Wojcieszak, Magdalena},
  journal={PNAS Nexus},
  volume={3},
  number={12},
  pages={pgae518},
  year={2024},
  doi={10.1093/pnasnexus/pgae518}
}

@article{milli2025engagement,
  title={Engagement, user satisfaction, and the amplification of divisive content on social media},
  author={Milli, Smitha and Carroll, Micah and Wang, Yike and Pandey, Sashrika and Zhao, Sebastian and Dragan, Anca D.},
  journal={PNAS Nexus},
  volume={4},
  number={3},
  pages={pgaf062},
  year={2025},
  doi={10.1093/pnasnexus/pgaf062}
}

@article{mattis2024nudging,
  title={Nudging towards news diversity: A theoretical framework for facilitating diverse news consumption through recommender design},
  author={Mattis, Nicolas and Masur, Philipp K. and M{\"o}ller, Judith and van Atteveldt, Wouter},
  journal={New Media \& Society},
  volume={26},
  number={7},
  year={2024},
  doi={10.1177/14614448221104413}
}

@article{wang2020public,
  title={Public Discourse and Social Network Echo Chambers Driven by Socio-Cognitive Biases},
  author={Wang, Xin and Sirianni, Antonio D. and Tang, Shaoting and Zheng, Zhiming and Fu, Feng},
  journal={Physical Review X},
  volume={10},
  number={4},
  pages={041042},
  year={2020},
  doi={10.1103/PhysRevX.10.041042}
}

@article{piao2023human,
  title={Human--AI adaptive dynamics drives the emergence of information cocoons},
  author={Piao, Jinghua and Liu, Jiazhen and Zhang, Fang and Su, Jun and Li, Yong},
  journal={Nature Machine Intelligence},
  volume={5},
  pages={1214--1224},
  year={2023},
  doi={10.1038/s42256-023-00731-4}
}

@article{haroon2023auditing,
  title={Auditing YouTube's recommendation system for ideologically congenial, extreme, and problematic recommendations},
  author={Haroon, Muhammad and Wojcieszak, Magdalena and Chhabra, Anshuman and Liu, Xin and Mohapatra, Prasant and Shafiq, Zubair},
  journal={Proceedings of the National Academy of Sciences},
  volume={120},
  number={50},
  pages={e2213020120},
  year={2023},
  doi={10.1073/pnas.2213020120}
}

@article{hosseinmardi2024causal,
  title={Causally estimating the effect of YouTube's recommender system using counterfactual bots},
  author={Hosseinmardi, Homa and Ghasemian, Amir and Rivera-Lanas, Miguel and Ribeiro, Manoel Horta and West, Robert and Watts, Duncan J.},
  journal={Proceedings of the National Academy of Sciences},
  volume={121},
  number={8},
  pages={e2313377121},
  year={2024},
  doi={10.1073/pnas.2313377121}
}

@article{schmidt2017anatomy,
  title={Anatomy of news consumption on Facebook},
  author={Schmidt, Ana Luc{\'i}a and Zollo, Fabiana and Del Vicario, Michela and Bessi, Alessandro and Scala, Antonio and Caldarelli, Guido and Stanley, H. Eugene and Quattrociocchi, Walter},
  journal={Proceedings of the National Academy of Sciences},
  volume={114},
  number={12},
  pages={3035--3039},
  year={2017},
  doi={10.1073/pnas.1617052114}
}

@article{cinelli2021echo,
  title={The echo chamber effect on social media},
  author={Cinelli, Matteo and De Francisci Morales, Gianmarco and Galeazzi, Alessandro and Quattrociocchi, Walter and Starnini, Michele},
  journal={Proceedings of the National Academy of Sciences},
  volume={118},
  number={9},
  pages={e2023301118},
  year={2021},
  doi={10.1073/pnas.2023301118}
}

@article{scharkow2020intermediaries,
  title={How social network sites and other online intermediaries increase exposure to news},
  author={Scharkow, Michael and Mangold, Frank and Stier, Sebastian and Breuer, Johannes},
  journal={Proceedings of the National Academy of Sciences},
  volume={117},
  number={6},
  pages={2761--2763},
  year={2020},
  doi={10.1073/pnas.1918279117}
}

@article{ibrahim2023youtube,
  title={YouTube's recommendation algorithm is left-leaning in the United States},
  author={Ibrahim, Hazem and AlDahoul, Nouar and Lee, Sangjin and Rahwan, Talal and Zaki, Yasir},
  journal={PNAS Nexus},
  volume={2},
  number={8},
  pages={pgad264},
  year={2023},
  doi={10.1093/pnasnexus/pgad264}
}

@article{santos2021link,
  title={Link recommendation algorithms and dynamics of polarization in online social networks},
  author={Santos, Fernando P. and Lelkes, Yphtach and Levin, Simon A.},
  journal={Proceedings of the National Academy of Sciences},
  volume={118},
  number={50},
  pages={e2102141118},
  year={2021},
  doi={10.1073/pnas.2102141118}
}

@article{glickman2025feedback,
  title={How human--AI feedback loops alter human perceptual, emotional and social judgements},
  author={Glickman, Moshe and Sharot, Tali},
  journal={Nature Human Behaviour},
  volume={9},
  number={2},
  pages={345--359},
  year={2025},
  doi={10.1038/s41562-024-02077-2}
}

@article{bhadani2022diversity,
  title={Political audience diversity and news reliability in algorithmic ranking},
  author={Bhadani, Saumya and Yamaya, Shun and Flammini, Alessandro and Menczer, Filippo and Ciampaglia, Giovanni Luca and Nyhan, Brendan},
  journal={Nature Human Behaviour},
  volume={6},
  number={4},
  pages={495--505},
  year={2022},
  doi={10.1038/s41562-021-01276-5}
}

@article{brady2026redesigning,
  title={Redesigning algorithms to intervene on social norm misperceptions during a national election},
  author={Brady, William J. and Doyle, Meriel and Elnakouri, Abdo and Finkel, Eli J. and Jackson, Joshua Conrad and Kteily, Nour and Parker, Victoria and Puryear, Curtis and Spelman, Trevor and Teeny, Jacob and Torres, Mark},
  journal={Nature},
  volume={655},
  number={8124},
  pages={942--956},
  year={2026},
  doi={10.1038/s41586-026-10536-1}
}

@article{battiston2020networks,
  title={Networks beyond pairwise interactions: Structure and dynamics},
  author={Battiston, Federico and Cencetti, Giulia and Iacopini, Iacopo and Latora, Vito and Lucas, Maxime and Patania, Alice and Young, Jean-Gabriel and Petri, Giovanni},
  journal={Physics Reports},
  volume={874},
  pages={1--92},
  year={2020},
  doi={10.1016/j.physrep.2020.05.004}
}

@article{benson2016higher,
  title={Higher-order organization of complex networks},
  author={Benson, Austin R. and Gleich, David F. and Leskovec, Jure},
  journal={Science},
  volume={353},
  number={6295},
  pages={163--166},
  year={2016},
  doi={10.1126/science.aad9029}
}

@article{bui2025algorithmic,
  title={Algorithmic discrimination: A grounded conceptualization},
  author={Bui, Matthew and McIlwain, Charlton and Olojo, Seyi and Chang, Ho-Chun Herbert},
  journal={Information, Communication \& Society},
  volume={29},
  number={2},
  pages={398--416},
  year={2026},
  doi={10.1080/1369118X.2025.2516544}
}

@article{changfang2024taiwan,
  title={The 2024 {Taiwanese} Presidential Election on social media: Identity, policy, and affective virality},
  author={Chang, Ho-Chun Herbert and Fang, Yu Sunny},
  journal={PNAS Nexus},
  volume={3},
  number={4},
  pages={pgae130},
  year={2024},
  doi={10.1093/pnasnexus/pgae130}
}

@article{iacopini2019simplicial,
  title={Simplicial models of social contagion},
  author={Iacopini, Iacopo and Petri, Giovanni and Barrat, Alain and Latora, Vito},
  journal={Nature Communications},
  volume={10},
  number={1},
  pages={2485},
  year={2019},
  doi={10.1038/s41467-019-10431-6}
}

@inproceedings{kempe2003maximizing,
  title={Maximizing the spread of influence through a social network},
  author={Kempe, David and Kleinberg, Jon and Tardos, {\'E}va},
  booktitle={Proceedings of the Ninth ACM SIGKDD International Conference on Knowledge Discovery and Data Mining},
  pages={137--146},
  year={2003},
  doi={10.1145/956750.956769}
}

@article{lerman2010information,
  title={Information contagion: An empirical study of the spread of news on {Digg} and {Twitter} social networks},
  author={Lerman, Kristina and Ghosh, Rumi},
  journal={Proceedings of the International AAAI Conference on Web and Social Media},
  volume={4},
  number={1},
  pages={90--97},
  year={2010},
  doi={10.1609/icwsm.v4i1.14021}
}

@inproceedings{nguyen2014exploring,
  title={Exploring the filter bubble: The effect of using recommender systems on content diversity},
  author={Nguyen, Tien T. and Hui, Pik-Mai and Harper, F. Maxwell and Terveen, Loren and Konstan, Joseph A.},
  booktitle={Proceedings of the 23rd International Conference on World Wide Web},
  pages={677--686},
  year={2014},
  doi={10.1145/2566486.2568012}
}

@techreport{page1999pagerank,
  title={The {PageRank} citation ranking: Bringing order to the web},
  author={Page, Lawrence and Brin, Sergey and Motwani, Rajeev and Winograd, Terry},
  institution={Stanford InfoLab},
  year={1999}
}

@article{stray2021designing,
  title={Designing recommender systems to depolarize},
  author={Stray, Jonathan},
  journal={First Monday},
  volume={27},
  number={5},
  year={2022},
  doi={10.5210/fm.v27i5.12604}
}

@article{watts2002simple,
  title={A simple model of global cascades on random networks},
  author={Watts, Duncan J.},
  journal={Proceedings of the National Academy of Sciences},
  volume={99},
  number={9},
  pages={5766--5771},
  year={2002},
  doi={10.1073/pnas.082090499}
}
\endgroup

\end{document}